Коротко о быстром поиске кратчайших путей в невзвешенном динамическом графе по его проекциям

В.А. Мелентьев

On the quick search for the shortest paths in an unweighted dynamic graph by its projections in brief

V.A. Melent'ev

Впервые предложены: способ представления проекций графа в памяти компьютера и основанное на нем описание быстрого поиска кратчайших путей в невзвешенных динамических графах. Пространственная сложность описания проекции не превышает $(d + 1) \cdot n$ слов, где $d$ — диаметр, а $n$ — число вершин графа. Временная сложность поиска одного кратчайшего пути между двумя вершинами не превышает $d$ шагов с длительностью элементарного времени выборки машинного слова. Решение может найти применение в критичных к временным задержкам протоколах маршрутизации компьютерных сетей и суперкомпьютеров.

Ключевые слова: невзвешенный граф, проекция графа, путевая таблица, кратчайший путь, временная и пространственная сложности поиска

Здесь мы ограничимся двумя вариантами невзвешенного графа: неориентированный (все ребра двунаправленны) и ориентированный граф с наличием однонаправленных дуг. В обоих вариантах следует найти путь минимальной длины между двумя заданными вершинами. Известные способы нахождения такого пути основаны, как правило, на двух традиционных способах представления графов в памяти компьютера: на наборе списков смежных вершин, дающем более компактное представление для разреженных (sparse) графов и на матрице смежности для плотных (dense) графов. Как для неориентированных так и для ориентированных графов представление графа $G(V, E)$ в виде списков требует объема памяти, равного $\Theta(n + m)$, где $n = |V|$, $m = |E|$. Матрице смежности графа требуется $\Theta(n^2)$ памяти независимо от количества его ребер и дуг [1]. Используемое здесь проективное описание впервые предложено и развито в работах [2-4], — там же можно достаточно подробно ознакомиться со свойствами проекций графа и их обоснованиями. Использование такого описания в топологическом моделировании параллельных систем [5], в оценках топологической совместимости параллельных задач и систем [6], в детерминированном синтезе релевантных заданным свойствам топологий вычислительных систем [7] дополняют его. Однако, ни в одной из этих работ до сих пор не рассматривались способы представление проекций в памяти компьютера. По-видимому, такие формы могут быть различными и зависят от решаемых с помощью проекций проблем.

Для решения проблемы быстрого поиска здесь предлагается описание проекции в виде таблицы с числом столбцов, равным числу вершин $n$, и числом строк, равным номеру уровня проекции $l_{full}$, на котором достигается ее вершинная полнота[1]. Уровни проекции последовательно пронумерованы снизу-вверх от нулевого, на котором расположена ракурсная вершина проекции, она же – ее основание, или вершина-источник $s$ (source), из которой необходимо найти кратчайшие пути в другую вершину $t$ (terminal) графа.

Заполнение таблицы (назовем ее путевой) начинаем со строки под номером, равным номеру верхнего уровня проекции, для этого в номера ячеек этой строки, соответствующие вершинам этого уровня, записываем номера порождающих их вершин нижерасположенного уровня. Аналогично поступаем с остальными строками таблицы и уровнями проекции вплоть до первой строки и первого уровня. Заполнение путевой таблицы можно проводить и в обратном порядке, начав с первой строки и первого уровня проекции и закончив последней строкой (последним уровнем проекции) — как будет удобно, разницы нет. Добавим к полученной таблице строку, обозначив ее для определенности как нулевую[2] (в таком случае ни одна из вершин графа не должна иметь такой номер), выделим ее серым цветом в знак того, что она в отличие от прочих строк несет в себе информацию только о длине пути. Заполним эту строку следующим образом: в ячейку под номером ракурсной вершины *s* (вершины-источника), запишем 0 — она расположена на нулевом уровне, и ее расстояние до самой себя равно нулю. В остальные ячейки этой строки поместим номера уровней, пересечение которых с соответствующими этим ячейкам столбцами не является пустым, уровни при этом записываем в порядке их возрастания. Тогда цифры в каждой ячейке нулевой строки указывают на длины путей от вершины-источника до вершины с номером этой ячейки.

---
[1] Проекция считается вершинно-полной, если в ней перечислены все вершины графа.
[2] Не отождествлять ее с нулевым уровнем проекции

Продемонстрируем вышесказанное примерами неориентированного и ориентированного графов, показанных на рисунках 1 и 2 и соответствующих им таблицах 1а, 1б, 2а, 2б. В целях лучшего восприятия для каждого графа приведены три проекции и соответствующие им таблицы.

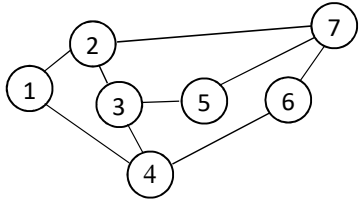

а) $P(1) = 1^{(2^{(3^{(4,5)}, 7^{(5,6)})}, 4^{(3^{(2,5)}, 6^{(7)})})}$,

б) $P(3) = 3^{(2^{(1,7)}, 4^{(1,6)}, 5^{(7)})}$,

в) $P(5) = 5^{(3^{(2^{(1,7)}, 4^{(1,6)})}, 7^{(2^{(1,3)}, 6^{(4)})})}$.

*Рис. 1* Неориентированный невзвешенный граф и его проекции (а, б) из вершин 1 и 3

*Таблица 1а* маршрутизации графа рис. 1 из вершины 1

|   | **1** | **2** | **3** | **4** | **5** | **6** | **7** |
|---|---|---|---|---|---|---|---|
| **3** |   | 3 |   | 3 | ③,7 | 7 | 6 |
| **2** |   |   | ②,4 |   |   | 4 | 2 |
| **1** |   | ① |   | 1 |   |   |   |
| **0** | 0 | 1,3 | 2 | 1,3 | ③ | 2,3 | 2,3 |

*Таблица 1б* маршрутизации графа рис. 1 из вершины 3

|   | **1** | **2** | **3** | **4** | **5** | **6** | **7** |
|---|---|---|---|---|---|---|---|
| **2** | 2,4 |   |   |   |   | 4 | 2,5 |
| **1** |   | 3 |   | 3 | 3 |   |   |
| **0** | 2 | 1 | **0** | 1 | 1 | 2 | 2 |

*Таблица 1в* маршрутизации графа рис. 1 из вершины 5

|   | **1** | **2** | **3** | **4** | **5** | **6** | **7** |
|---|---|---|---|---|---|---|---|
| **3** | ②,4 |   | 2 | 6 |   | 4 | 2 |
| **2** |   | ③,7 |   | 3 |   | 7 |   |
| **1** |   |   | ⑤ |   |   |   | 5 |
| **0** | ③ | 2 | 1,3 | 2,3 | **0** | 2,3 | 1,3 |

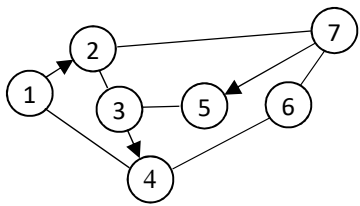

а) $P(1) = 1^{(2^{(3^{(4,5)}, 7^{(5,6)})}, 4^{(6^{(7)})})}$,

б) $P(3) = 3^{(2^{(7)}, 4^{(1,6)}, 5)}$,

в) $P(5) = 5^{(3^{(2^{(7)}, 4^{(1,6)})})}$.

*Рис. 2* Ориентированный невзвешенный граф и его проекции (а, б) из вершин 1 и 3

*Таблица 2а* маршрутизации графа рис. 2 из вершины 1

|   | **1** | **2** | **3** | **4** | **5** | **6** | **7** |
|---|---|---|---|---|---|---|---|
| **3** |   |   |   | 3 | ③,7 | 7 | 6 |
| **2** |   |   | ② |   |   | 4 | 2 |
| **1** |   | ① |   | 1 |   |   |   |
| **0** | 0 | 1 | 2 | 1,3 | ③ | 2,3 | 2,3 |

*Таблица 2б* маршрутизации графа рис. 2 из вершины 3

|   | **1** | **2** | **3** | **4** | **5** | **6** | **7** |
|---|---|---|---|---|---|---|---|
| **2** | 4 |   |   |   |   | 4 | 2 |
| **1** |   | 3 |   | 3 | 3 |   |   |
| **0** | 2 | 1 | **0** | 1 | 1 | 2 | 2 |

*Таблица 2*в маршрутизации графа рис. 2 из вершины 5

|   | 1 | 2 | 3 | 4 | 5 | 6 | 7 |
|---|---|---|---|---|---|---|---|
| **3** | ④ |   |   |   |   | 4 | 2 |
| **2** |   | 3 |   | ③ |   |   |   |
| **1** |   |   |   | ⑤ |   |   |   |
| **0** | ③ | 2 | 1 | 2 | **0** | 3 | 3 |

Использование полученных из проекций графа таблиц позволяет достаточно просто выбрать кратчайшие из заданных этими таблицами путей, причем наличие или отсутствие в графе ориентированных ребер (дуг) не меняет процедуру их получения.

Итак, необходимо получить кратчайший путь из вершины *s* в вершину *t* в некотором невзвешенном динамическом графе $G(V,E)$, его проекции $P(v_i)$ и диаметр – $d(G)$ соответствуют состоянию системы на момент поиска. В [8] можно ознакомиться с формальными процедурами актуализации проекций графа.

Как показано выше, полученная из проекции $P(s)$ путевая таблица содержит $d + 1$ строк и *n* столбцов, т. е пространственная сложность поиска кратчайшего пути из вершины *s* в любую другую вершину графа не превысит $O((d + 1)\,n)$. Заметим, что кратчайших путей может быть несколько и для поиска одного из них достаточно всего одного элемента в определяющих этот путь ячейках, однако в целях сохранения возможности адаптивного поиска путей, учитывающего динамику изменения состояния системы, описываемой исследуемым графом, считаем, что ячейки могут содержать в себе динамически упорядоченные по приоритету множества элементов, что позволит выбирать один из наиболее актуальных на текущий момент кратчайших путей. Если же ячейка таблицы пуста, то соответствующая ей вершина не входит в состав множества искомых кратчайших путей.

Суть поиска состоит в следующем: первый элемент *t*-й ячейки нулевой строки определяет длину $l(s,t)$ кратчайшего пути и равен номеру наименьшего уровня проекции $P(s)$, содержащего вершину *t*. Выборка первого элемента из *t*-й ячейки *l*-й строки определяет вершину $v_{l-1}$, смежную вершине *t* и расположенную от вершины *s* на расстоянии $l - 1$. Аналогично определяются и все остальные вершины искомого пути вплоть до строки $l = 2$, пересечение которой со столбцом, соответствующим определенной на предыдущем шаге вершине, неизбежно укажет на вершину *s*. Т. е. число таких шагов равно уменьшенной на единицу длине кратчайшего пути. Добавив к этому числу шаг выборки длины пути из нулевой строки таблицы, получим, что временная сложность поиска одного кратчайшего пути между двумя вершинами в графе не превысит его диаметра $O(d)$, и учитывая, что каждый из *d* шагов, по сути, элементарен и заключается лишь в выборке содержимого ячейки путевой таблицы, процесс этот быстротечен.

Изложенная выше последовательность действий по определению путей $R(s,t)$ из вершины 1 в вершину 5 и из 5 в 1 продемонстрирована в таблицах 1а и 1в — для неориентированного (рис. 1) графа, и в таблицах 2а и 2в — для ориентированного (рис. 2) графа. Штриховая линии и ее направление указывают на начальную и конечную вершины искомого кратчайшего пути. Для рис. 1 получены пути $R(1,5) = (1, 2, 3, 5)$ и $R(5,1) = (5, 3, 2, 1)$, причем прямой путь $R(1,5)$ здесь равен обратному $R^{-1}(5,1)$. В ориентированном графе рис. 2 такого совпадения нет: $R(1,5) = (1, 2, 3, 5)$, $R(5,1) = (5, 3, 4, 1)$, и $R(1,5) \neq R^{-1}(5,1)$.

Сочетание презентованного здесь представления проекций графа в памяти, например, сетевых коммутаторов (маршрутизаторов), основанного на этом представлении поиска, и элементарность выполняемых при поиске операций должно существенно снизить латентность интерконнекта компьютерных сетей и ссуперкомпьютеров. Развернутое изложение представленного материала будет включать в себя и взвешенные динамические графы.